# Towards effective research recommender systems for repositories

Petr Knoth, Lucas Anastasiou, Aristotelis Charalampous, Matteo Cancellieri, Samuel Pearce, Nancy Pontika, Vaclav Bayer

CORE, Knowledge Media institute, The Open University

{petr.knoth, lucas.anastasiou, aristotelis.charalampous, matteo.cancellieri, samuel.pearce, nancy.pontika, vaclav.bayer}@open.ac.uk

**Session Type**

- Presentation

**Abstract**

In this paper, we argue why and how the integration of recommender systems for research can enhance the functionality and user experience in repositories. We present the latest technical innovations in the CORE Recommender, which provides research article recommendations across the global network of repositories and journals. The CORE Recommender has been recently redeveloped and released into production in the CORE system and has also been deployed in several third-party repositories. We explain the design choices of this unique system and the evaluation processes we have in place to continue raising the quality of the provided recommendations. By drawing on our experience, we discuss the main challenges in offering a state-of-the-art recommender solution for repositories. We highlight two of the key limitations of the current repository infrastructure with respect to developing research recommender systems: 1) the lack of a standardised protocol and capabilities for exposing anonymised user-interaction logs, which represent critically important input data for recommender systems based on collaborative filtering and 2) the lack of a voluntary global sign-on capability in repositories, which would enable the creation of personalised recommendation and notification solutions based on past user interactions.

**Conference Themes**

*List the conference theme(s) your proposal best addresses (remove the others):*

- Supporting Open Scholarship, Open Data, and Open Science
- Repositories of high volume and/or complex data and collections
- Integrating with the Wider Web and External Systems

**Audience**

Repository managers, developers, librarians, researchers

**Background**

The submission addresses the challenges in the design and the development of research recommender systems on top of the global network of open repositories. We discuss and make recommendations for technical innovations in the repositories infrastructure (discussed also as part of the COAR Next Generation Repositories Group), such as a voluntary global sign-on and functionality for openly releasing anonymised user-interaction data, to enable the creation of effective research recommender systems in the future.

# Introduction

While repositories can be seen as fairly successful tools for the preservation of research papers, they have been often criticised for the limited *accessibility* (sometimes also referred to as *discoverability*) of resources they offer access to (Salo, 2008; Konkiel, 2012).

Accessibility is an abstract concept originally used in urbanism (Hansen, 1959) defined as a measure of potential opportunities for interaction with resources, such as employment, schooling or dining. Azzopardi and Vinay (2008) use an analogy of this definition to define accessibility in the context of information retrieval as the potential of documents for retrieval. Based on this definition, we can say, for example, that a Web a page with no incoming links will have intuitively lower accessibility than a page with many incoming links, as there will be fewer opportunities to visit it.

One of the main purposes of repositories is to disseminate (open access) research outputs across the world. Consequently, it is important that these outputs are well accessible, i.e. that they are sufficiently connected to provide enough opportunities for interaction with them.

There are many different types of search behavior and accessibility needs to be evaluated with respect to each use case separately. For traditional lookup search, which can be defined as retrieving an item with respect to a user-defined query, we can use the approach of Azzopardi and Vinay (2008) to measure accessibility of a resource based on the likelihood of expressing a query leading to the retrieval of the resource and the position of the resource in the search results list. Repositories have been for a long time criticised for poor accessibility of their content in lookup search (Salo, 2008; Konkiel, 2012), such as in Google, and recommendations how to mitigate these problems have analysed and discussed (Acharya, 2015).

However, researchers are often immersed in more complex search activities, the so-called exploratory search (Marchionini, 2006; Herrmannova & Knoth, 2012). This type of search is characteristic by the difficulty of formulating effective queries and involves, for example, analysis, comparison, exclusion/negation and synthesis. Recommender systems have been widely used as an effective answer to exploratory (or browsing) activities.

We claim that by integrating recommender systems into repositories, we can increase the number of incoming links to relevant research outputs, thereby increase the accessibility of these resources for exploratory search activities.

# Research recommender systems and repositories

Recommender systems are increasingly used on the Web to help users find material relevant to their interests. While they have been highly successful in commercial environments to boost sales, there exist only few recommenders for research. To our knowledge, we have been the first to explore the development of recommenders based on and for repositories.

From a user's perspective, recommenders are either personalised, i.e. recommendations are based on the knowledge of the user's preferences or past activity, or non-personalised, recommendations are the same for all users. There are two important classes of recommenders: collaborative filtering (CF) and content-based filtering (CBF).

CF makes use of past interaction data to recommend new unknown items to a user based on the assumption of similarity (e.g. to other users in the case of user-based CF). These techniques have proven extremely effective in supporting exploratory browsing. However, in order to work effectively, they require a sufficiently large amount of user-generated data. In addition, CF cannot recommend items no user has acted upon yet, the so-called cold items. Therefore, the strategy of many recommender systems is to find an unintrusive way to expose cold items to users, for example, by blending them discretely to a homepage where users are more likely to click on them, or by applying content-based filtering, effectively encouraging the decrease of cold items in the database.

While CF-based recommenders can be seen as the current state-of-the-art in recommender systems, it is challenging to apply them in open repositories. In order to achieve that, repositories would need to (ideally on a global scale) expose user-interaction data, so that these data can be utilised for building cross-repository recommender systems for research. While we and others (Wesely-Smith et al., 2015) believe that there is a reasonable case for this, it is without doubts a capability/functionality which many might reject for privacy reasons.

The alternative approach to build recommender systems for repositories is to rely on CBF. This method attempts to recommend items using the assumption of content similarity, such as cosine, based on attributes (features) of each item. In case of research papers, these are, for example, title, abstract, field of study, journal name, affiliation, publication year, number of readers/downloads and full text. The main advantages of CBF are that it does not suffer from the cold-start problem as CF and can still be applied to produce both personalised and non-personalised recommendations.

# The CORE recommender system

The first version of the CORE Recommender was developed in 2011. In 2016, we redeveloped the system and made many improvements. The new recommender was released in production in autumn 2016 and apart from the fact that it powers related article recommendations in CORE, it is also deployed at several repositories and journals.

*Figure 1: The CORE recommender plugin as embedded in the WhiteRose Research Online Repository. Tabs are used to provide both suggestions from within the repository as well as across all repositories.*

The integration of the recommender in EPrints repositories is especially easy as we provide it as an EPrints Bazaar Plugin, but a solution for DSPace, OJS and virtually any webpage is also available. When a user views an article page in a repository, the plugin sends to CORE information about the visited item. This can include the item's identifier and, when possible, also its metadata. CORE then replies back to the repository system embedding a list of suggested articles for reading. These are currently presented on two tabs, suggested articles in other repositories and suggested articles in this repository.

## How does the CORE recommender algorithm work?

As of November 2016, there are more than 37.5 million metadata records and 4.5 million full texts from a global network of 2,300 repositories and 6,000 journals in CORE. This makes CORE, to our knowledge, the world's largest full text aggregator of open access content. Access to this multidisciplinary dataset enables the CORE recommender, which is a non-personalised recommender based on CBF, to produce relevant recommendations given the vast majority of possible queries (which are expressed as reference research papers).

The recommendation algorithm takes as an input a reference document, which it represents as a set of features using the Vector Space Model. These features include the title, authors, abstract and publication year. If the reference document can be at query time matched to an existing document in CORE, we also make use of the full text and possibly richer metadata. The latter can be provided either explicitly by one of the potentially many repositories from which the document was harvested during the aggregation or can be inferred by one of our content enrichment algorithms. The inferred metadata can include, for instance, the language of the source and key terms. Last but not least, we rely on identifiers, such as the DOI, to enrich the feature space with citation counts, downloads and Mendeley readership indicators by relying on 3rd party systems and datasets. All these features are then used to find the closest matches in the CORE corpus at query time (caching is applied).

Of course, not every feature has the same importance. In our internal ranking algorithm we positively or negatively boost some features to produce a better list of top recommendations. Numeric features are treated differently though. In the case of publication year, we apply a standard exponential decay function to promote more recent recommendations.

**Evaluating the quality of recommendations**

There is a range of parameters influencing the quality of the recommendation output, such as of the time decay function or the boosting applied to features representing a particular metadata field. Our approach to set these parameters is data-driven, which requires us to run experiments evaluating the performance of each parameter combination to find a suitable set-up. The performance is measured in two scenarios: *offline* and *online*.

For the offline scenario, we have created a ground-truth based on 500 million+ citation relationships from the Microsoft Academic Graph dataset (Arnab et al., 2015). For the offline evaluation purposes, we use the "citation" assumption that "if an article X cites article Y, then recommending article X given article Y and vice versa is considered as the correct answer"[1]. We benchmarked our recommender for various parameter configurations to reach higher performance as measured by standard metrics used in recommender systems, namely precision, recall, mean average precision, NDCG and others (Manning et al., 2008).

However, the "citation" assumption is just one possible imperfect and sparse proxy to an ideal ground-truth. To increase our trust in the experimental results and to understand robustness of each parameter set up, we have also developed another ground-truth based on the "co-citation" assumption. Under this assumption, "if an article X is often co-cited with an article Y in research papers, then recommending article X given article Y is considered as the correct answer". This assumption can be further extended by taking into account the positionality/proximity of the citations within the document (Gipp & Beel, 2009).

Offline evaluation in recommender systems is often not sufficiently predictive of the online experience (Beel & Langer, 2015). This means that a good offline performance of an algorithm does not reliably predict a good online performance of the same algorithm, when deployed in a real production system. Consequently, we need to carry on with the testing and fine-tuning of the performance on a continuous basis. To address this problem, we measure the Click Through Rate (CTR), which is a well-established on-line evaluation metric used widely by recommender systems, and we regularly revisit the scores. Before introducing new algorithms into production or changing parameters, it is useful to perform A/B testing (Kohavi & Longbotham, 2015), ensuring that the CTR score of the new setup is statistically significantly better that the CTR of the existing setup.

**Post-filtering and crowd-sourcing as an error prevention mechanism**

The CORE recommender employs several filters as a mechanism to increase the quality of the generated recommendation lists. We only recommend: papers with an open access full-text; articles with at least a minimal set of metadata attributes; articles with a generated thumbnail; articles for which we haven't received negative feedback, etc. We also deduplicate the resulting set of recommendations ensuring they refer to different documents.

Despite these efforts, we are aware that, in some cases, the recommender might still provide irrelevant recommendations. One of the ways in which we try to decrease the frequency of these events is by offering our users the ability to report irrelevant recommendations using a feedback button. These items are then blacklisted and will not be displayed again in the recommendation list. In a sense, we are crowd-sourcing a recommendations blacklist to improve the performance.

**Future work**

At the moment, the CORE recommender provides non-personalised content-based article recommendations from across the global open access repositories and journals network. In the future, we want to transform our recommender from a pure CBF recommendation engine to a solution that can also deliver personalised recommendations using CF based on past user behaviour. This requires the creation of user profiles and consequently faces many challenges. However, once this is done, it can pave the way for the delivery of research notification services developed on top of the repositories' infrastructure - similarly as we might know them from some research social networks - but with a high degree of data transparency and reliance on open data.

Additionally, we want to start recommending not just articles, but other entities participating in the research process, especially research funding opportunities, collaborators/experts and research events (conferences, workshops, journals). Finally, there are many more exciting possibilities, such as recommending research methods and auto-suggesting citations to back-up claims in real time while writing research literature.

# Discussion

Despite the work done by CORE, developing recommender systems for repositories faces many challenges. The first challenge is that for a recommender system to work, a fast access to a global pool of research literature is required.

---

[1] The recommender currently does not use the structure of the citation network to decide what to recommend, however, it does have access to aggregated citation counts.

Harvesting content from repositories is a fairly complex task itself (Knoth, 2013). Aggregators are therefore a natural choice for providing these solutions.

The second challenge is the difficulty in providing personalised recommenders for repositories. Personalised recommenders require access to a user profile, which is either explicitly or implicitly built by the user. However, repositories have been originally set up so that a user sign on is not required, and this has obviously been done for very good reasons. The key question is therefore how to keep repositories open to unauthorised users while providing personalisation support to those users who would like to opt-in for such a service.

We believe that there would be little value for each repository to implement a local user registration. Voluntary authorisation can only make a difference provided that the user is recognised across a global network of repositories. We think that there are two services that can potentially be useful in this context. In theory, if researchers could sign on in any repository via an ORCID iD, and then be recognised across any other repository they visit, it would be possible to provide personalised solutions based on all data provided in their ORCID profiles. This includes especially the list of the author's publications, i.e. a very useful resource for personalisation indeed. However, not all repository users are researchers and consequently, using, for example, a Twitter sign on, it might be possible to understand, in some cases, the person's research interests. While it is hard to imagine that the voluntary sign on functionality would be implemented by repositories and adopted by users just to provide a solution for recommender systems, it could be highly valuable for many other services that can be built on top of repositories too, such as commenting and globally enabled depositing of research papers.

But one important issue still remains. Unless anonymised user interaction data (for example, in the form of an anonymised user id, document id and access time) are exposed by repositories and ideally made available for harvesting globally, it will continue to be challenging for most 3rd party services to develop recommender systems for repositories based on CF. While there is always the alternative of CBF, CF recommenders are currently known to produce better results in most scenarios. An important issue is that CF recommender systems for research can already be developed by companies that have many users on their platform and consequently collect and own such rich interaction data. However, these solutions are unlikely to ever become fully transparent. There is a real danger that unless equal access to such data is provided publicly by the global repositories network, the reliance of universities on these commercial providers will further increase, as more researchers become dependent on their services in their daily activities leading to a vicious cycle. Consequently, we believe that although such steps might seem unpopular, it might actually be in the interest of the repositories' community to consider moving in the direction of openly and transparently exposing such interaction data.

The two important proposed capabilities for the repositories network, namely 1) adding support for recommenders systems by exposing interaction data and 2) a voluntary global sign on are currently being discussed as part of the COAR Next Generation Repositories working group (COAR, 2016).

## Conclusions

Despite many technical challenges, the application of recommender systems in research have a lot of potential for improving the way researchers are kept up to date and collaborate. Recommender system can help users find, explore and discover research as soon as it becomes available and can support exploratory serendipitous browsing. There haven't been many recommenders deployed in the academic domain yet, but it is certainly a growing field.

The CORE Recommender is unique in a number of aspects. Firstly, our methods rely on the availability of full texts and are not based on abstracts or metadata only. Secondly, we ensure that the recommended articles are available as open access, making the CORE Recommender a good choice not only for open repositories, but also for the general public. Thirdly, to our knowledge this is the only recommender developed on top of the repositories' network content and intended for repositories. We provide our recommendation service for free and it is also available using our API. We believe that the CORE Recommender is an example of the added value that can be delivered over the global network of repositories.

Finally, we suggest to find a solution for the repositories infrastructure to 1) expose anonymised user-interaction data and 2) offer a voluntary global sign on, as this would expand the functionality that can currently be provided recommender systems for research and would take the performance of the solutions to a new level.